# Polynomial-time Computation of Exact Correlated Equilibrium in Compact Games

Albert Xin Jiang and Kevin Leyton-Brown
Dept. of Computer Science, University of British Columbia
{jiang; kevinlb}@cs.ubc.ca

October 31, 2018

**Abstract**

In a landmark paper, Papadimitriou and Roughgarden [2008] described a polynomial-time algorithm ("Ellipsoid Against Hope") for computing sample correlated equilibria of concisely-represented games. Recently, Stein, Parrilo and Ozdaglar [2010] showed that this algorithm can fail to find an exact correlated equilibrium, but can be easily modified to efficiently compute approximate correlated equilibria. Currently, it remains unresolved whether the algorithm can be modified to compute an exact correlated equilibrium. We show that it can, presenting a variant of the Ellipsoid Against Hope algorithm that guarantees the polynomial-time identification of exact correlated equilibrium. Our new algorithm differs from the original primarily in its use of a separation oracle that produces cuts corresponding to pure-strategy profiles. As a result, we no longer face the numerical precision issues encountered by the original approach, and both the resulting algorithm and its analysis are considerably simplified. Our new separation oracle can be understood as a derandomization of Papadimitriou and Roughgarden's original separation oracle via the method of conditional probabilities. Also, the equilibria returned by our algorithm are distributions with polynomial-sized supports, which are simpler (in the sense of being representable in fewer bits) than the mixtures of product distributions produced previously; no tractable algorithm has previously been proposed for identifying such equilibria.

## 1 Introduction

We consider the problem of computing a sample correlated equilibrium [Aumann, 1974; Aumann, 1987] given a finite, simultaneous-move game. It is known that correlated equilibria of a game can be formulated as probability distributions over pure strategy profiles satisfying certain linear constraints. The resulting linear feasibility program has size polynomial in the size of the normal form representation of the game. However, the size of the normal form representation grows exponentially in the number of players. This is problematic when games involve large numbers of players. Fortunately, most large games of practical interest have highly-structured payoff functions, and thus it is possible to represent them compactly. A line of research thus exists to look for *compact game representations* that are able to succinctly describe structured games, including work on graphical games [Kearns *et al.*, 2001] and action-graph games [Bhat & Leyton-Brown, 2004; Jiang *et al.*, 2010]. But now the size of the linear feasibility program for CE can be exponential in the size of compact representation; furthermore a CE can require exponential space to specify.

The "Ellipsoid Against Hope" algorithm [Papadimitriou, 2005; Papadimitriou & Roughgarden, 2008] is a polynomial-time method for identifying a (polynomial-size representation of a) CE, given a game representation with *polynomial type* and satisfying the *polynomial expectation property*. Many existing compact game representations (including graphical games, symmetric games, congestion games, polymatrix games and action-graph games) satisfy these properties. This important result extends CE's attractive computational properties to the case of compactly represented games; note in contrast that the problem of finding a Nash equilibrium is PPAD-hard for normal form games [Chen & Deng, 2006; Daskalakis & Papadimitriou, 2005] as well as for certain compact game representations [Goldberg & Papadimitriou, 2006; Jiang *et al.*, 2010].

At a high level, the Ellipsoid Against Hope algorithm works by solving an infeasible dual LP ($D$) using the ellipsoid method (exploiting the existence of a separation oracle), and arguing that the LP ($D'$) formed by the generated cutting planes must also be infeasible. Solving the dual of this latter LP (which has polynomial size) yields a CE, which is represented as a mixture of the product distributions generated by the separation oracle.

## 1.1 Recent uncertainty about the complexity of exact CE

In a recent paper, Stein, Parrilo and Ozdaglar [2010] raised two interrelated concerns about the Ellipsoid Against Hope algorithm. First, they identified a symmetric 3-player, 2-action game with rational utilities on which the algorithm can fail to compute an exact CE. Indeed, they showed that the same problem arises on this game for a whole class of related algorithms. Specifically, if an algorithm (a) outputs a rational solution, (b) outputs a convex combination of product distributions, and (c) outputs a convex combination of symmetric product distributions when the game is symmetric, then that algorithm must fail to find an exact CE on their game, because the only CE of their game that satisfies properties (b) and (c) has irrational probabilities. This implies that any algorithm for exact rational CE must violate either (b) or (c).

Second, Stein, Parrilo and Ozdaglar also showed that the original analysis in [Papadimitriou & Roughgarden, 2008] incorrectly handles certain numerical precision issues, which we now briefly describe. Recall that a run of the ellipsoid method requires as inputs an initial bounding ball with radius $R$ and a volume bound $v$ such that the algorithm stops when the ellipsoid's volume is smaller than $v$. To correctly certify the (in)feasibility of an LP using the ellipsoid method, $R$ and $v$ need to be set to appropriate values, which depend on the maximum encoding size of a constraint in the LP. However (as pointed out by Papadimitriou and Roughgarden [2008]), each cut returned by the separation oracle is a convex combination of the constraints of the original dual LP $(D)$ and thus may require more bits to represent than any of the constraints in $(D)$; as a result, the infeasibility of the LP $(D')$ formed by these cuts is not guaranteed. Papadimitriou and Roughgarden [2008] proposed a method to overcome this difficulty, but Stein *et al.* show that this method is insufficient for finding an exact CE. For the related problem of finding an approximate correlated equilibrium ($\epsilon$-CE), Stein *et al.* give a slightly modified version of the Ellipsoid Against Hope algorithm that runs in time polynomial in $\log \frac{1}{\epsilon}$ and the game representation size. For problems that can have necessarily irrational solutions, it is standard to consider such approximations as efficient; however, there always exists a rational CE in a game with rational utilities, since CE are defined by linear constraints. It remains unresolved whether the Ellipsoid Against Hope algorithm can be modified to compute an exact, rational correlated equilibrium.[1]

## 1.2 Our results

In this paper, we use an alternate approach—completely sidestepping the issues just discussed—to derive a polynomial-time algorithm for computing an exact (and rational) correlated equilibrium given a game representation that has polynomial type and polynomial expectation property. Our algorithm has features that give it theoretical and practical value independently of the severity of the recently-identified issues in the original Ellipsoid Against Hope algorithm. Specifically, our approach is based on the observation that if we use a separation oracle (for the same dual LP formulation as in [Papadimitriou & Roughgarden, 2008]) that generates cuts corresponding to pure-strategy profiles (instead of Papadimitriou and Roughgarden's separation oracle that generates nontrivial product distributions), then these cuts are actual constraints in the dual LP, as opposed to the convex combinations of constraints produced by Papadimitriou and Roughgarden's separation oracle. As a result we no longer encounter the numerical accuracy issues that prevented the previous approaches from finding exact correlated equilibria. Both the resulting algorithm and its analysis are also considerably simpler than the original: standard techniques from the theory of the ellipsoid method are sufficient to show that our algorithm computes an exact CE using a polynomial number of oracle queries.

The key issue is the identification of pure-strategy-profile cuts. It is relatively straightforward to show that such cuts always exist: since the product distribution generated by the Ellipsoid Against Hope algorithm ensures the non-negativity of a certain expected value, then by a simple application of the probabilistic method there must exist a pure-strategy profile that also ensures the nonnegativity of that expected value. The key is to go beyond this nonconstructive proof of existence to also *compute* pure-strategy-profile cuts in polynomial time. We show how to do this by applying the method of conditional probabilities [Erdos & Selfridge, 1973; Spencer, 1994; Raghavan, 1988], an approach for derandomizing probabilistic proofs of existence. At a high level, our new separation oracle begins with the product distribution generated by Papadimitriou and Roughgarden's separation oracle, then sequentially fixes a pure strategy for each player in a way that guarantees that the corresponding conditional expectation given the choices so far remains nonnegative. Since our separation oracle goes though players sequentially, the cuts generated can be asymmetric even for symmetric games. Indeed, we can confirm (see Section 4.2) that it makes such asymmetric cuts on Stein, Parrilo and Ozdaglar's symmetric game—thus violating their condition (c)—because our algorithm always identifies a rational CE.

---

[1]In a recent online comment, Papadimitriou [2010] acknowledged the flaw in the original Ellipsoid Against Hope algorithm, but expressed confidence that the numerical issues can be overcome without dramatic changes (e.g., by using the repeated projection method due to Karp and Papadimitriou [1980]), yielding a polynomial algorithm for exact CE. We note also that, apparently believing that the issues they raised are not serious, Stein *et al.* have very recently withdrawn their paper from arXiv. It is our impression that their results are nevertheless still believed to be correct; we discuss them here because they help to motivate our alternate approach.

Another effect of our use of pure-strategy-profile cuts is that the correlated equilibria generated by our algorithm are guaranteed to have polynomial-sized supports; i.e., they are mixtures over a polynomial number of pure strategy profiles. Correlated equilibria with polynomial-sized supports are known to exist in every game (e.g., [Germano & Lugosi, 2007]) but no tractable algorithm has previously been proposed for identifying them. Such small-support correlated equilibria have a simpler form than the mixtures of product distributions produced by the Ellipsoid Against Hope algorithm, and may have certain practical advantages over the latter including requiring fewer bits to represent, being easier to sample from, and being easier to verify.

The rest of the paper is organized as follows. We start with basic definitions and notation in Section 2. In Section 3 we summarize Papadimitriou and Roughgarden's Ellipsoid Against Hope algorithm. In Section 4 we describe our algorithm and prove its correctness, and Section 5 concludes.

## 2 Preliminaries

We largely follow the notation of Papadimitriou [2005] and Papadimitriou and Roughgarden [2008]. Consider a simultaneous-move game with $n$ players. Denote a player $p$, and player $p$'s set of pure strategies (i.e., actions) $S_p$. Let $m = \max_p |S_p|$. Denote a pure strategy profile $s = (s_1, \ldots, s_n) \in S$, with $s_p$ being player $p$'s pure strategy. Denote by $S_{-p}$ the set of partial pure strategy profiles of the players other than $p$. Player $p$'s utility under pure strategy profile $s$ is $u_s^p$. We assume that utilities are nonnegative integers (but results in this paper can be straightforwardly adapted to the case of rational utilities). Denote the largest utility of the game as $u$.

A *correlated distribution* is a probability distribution over pure strategy profiles, represented by a vector $x \in \mathbb{R}^M$, where $M = \prod_p |S_p|$. Then $x_s$ is the probability of pure strategy profile $s$ under the distribution $x$. A correlated distribution $x$ is a *product distribution* when it can be achieved by each player $p$ randomizing independently over her actions according to some distribution $x^p$, i.e., $x_s = \prod_p x^p_{s_p}$. Such a product distribution is also known as a mixed-strategy profile, with each player $p$ playing the mixed strategy $x^p$.

Throughout the paper we assume that a game is given in a representation satisfying two properties, following Papadimitriou and Roughgarden [2008]:

- *polynomial type*: the number of players and the number of actions for each player are bounded by polynomials in the size of the representation.

- *polynomial expectation property*: We have access to an algorithm that computes the expected utility of any player $p$ under any product distribution $x$, i.e., $\sum_{s \in S} u^p_s x_s$, in time polynomial in the size of the representation.

**Definition 2.1.** *A correlated distribution $x$ is a* correlated equilibrium (CE) *if it satisfies the following* incentive constraints: *for each player $p$ and each pair of her actions $i, j \in S_p$,*

$$\sum_{s \in S_{-p}} [u^p_{is} - u^p_{js}] x_{is} \geq 0, \tag{1}$$

*where the subscript "is" (respectively "js") denotes the pure strategy profile in which player $p$ plays $i$ (respectively $j$) and the other players play according to the partial profile $s \in S_{-p}$.*

We write these incentive constraints in matrix form as $Ux \geq 0$. Thus $U$ is an $N \times M$ matrix, where $N = \sum_p |S_p|^2$. The rows of $U$, corresponding to the left-hand sides of the constraints (1), are indexed by $(p, i, j)$ where $p$ is a player and $i, j \in S_p$ are a pair of $p$'s actions. Denote by $U_s$ the column of $U$ corresponding to pure strategy profile $s$. These incentive constraints, together with the constraints

$$x \geq 0, \quad \sum_{s \in S} x_s = 1, \tag{2}$$

which ensure that $x$ is a probability distribution, form a linear feasibility program that defines the set of CE. The largest value in $U$ is at most $u$.

We define the *support* of a correlated equilibrium $x$ as the set of pure strategy profiles assigned positive probability by $x$. Germano and Lugosi [2007] showed that for any $n$-player game, there always exists a correlated equilibrium with support size at most $1 + \sum_p |S_p|(|S_p| - 1) = N + 1 - \sum_p |S_p|$. Intuitively, such correlated equilibria are basic feasible solutions of the linear feasibility program for CE, i.e., they are vertices of the polyhedron defining the feasible region. Furthermore, these basic feasible solutions involve only rational numbers for games with rational payoffs (see e.g. Lemma 6.2.4 of [Grötschel et al., 1988]).

## 3 The Ellipsoid Against Hope Algorithm

In this section, we summarize Papadimitriou and Roughgarden's Ellipsoid Against Hope algorithm for finding a sample CE [Papadimitriou, 2005; Papadimitriou

& Roughgarden, 2008], which can be seen as an efficiently constructive version of earlier proofs [Hart & Schmeidler, 1989; Nau & McCardle, 1990; Myerson, 1997] of the existence of CE. We will concentrate on the main algorithm and only briefly point out the numerical issues discussed at length by both Papadimitriou and Roughgarden [2008] and Stein *et al.* [2010], as our analysis will ultimately sidestep these issues.

Papadimitriou and Roughgarden's approach considers the linear program

$$\max \sum_{s \in S} x_s \qquad (P)$$
$$Ux \geq 0, \qquad x \geq 0,$$

which is modified from the linear feasibility program for CE by replacing the constraint $\sum_{s \in S} x_s = 1$ from (2) with the maximization objective. $(P)$ either has $x = 0$ as its optimal solution or is unbounded; in the latter case, the game has a correlated equilibrium. Thus one way to prove the existence of CE is to show the infeasibility of the dual problem

$$U^T y \leq -1, \qquad y \geq 0. \qquad (D)$$

The Ellipsoid Against Hope algorithm uses the following lemma, versions of which were also used by Nau and McCardle [1990] and Myerson [1997].

**Lemma 3.1** ([Papadimitriou, 2005; Papadimitriou & Roughgarden, 2008]). *For every dual vector $y \geq 0$, there exists a product distribution $x$ such that $xU^T y = 0$. Furthermore there exists an algorithm that given any $y \geq 0$, computes the corresponding $x$ (represented by $x^1, \ldots, x^n$) in time polynomial in $n$ and $m$.*

We will not discuss the details of this algorithm; we will only need the facts that the resulting $x$ is a product distribution and can be computed in polynomial time. Note also that the resulting $x$ is symmetric across players if $y$ is symmetric across players. Lemma 3.1 implies that the dual problem $(D)$ is infeasible (and therefore a CE must exist): $xU^T y$ is a convex combination of the left hand sides of the rows of the dual, and for any feasible $y$ the result must be less than or equal to $-1$.

The Ellipsoid Against Hope algorithm runs the ellipsoid algorithm on the dual $(D)$, with the algorithm from Lemma 3.1 as separation oracle, which we call the the Product Separation Oracle. At each step of the ellipsoid algorithm, the separation oracle is given a dual vector $y^{(i)}$. The oracle then generates the corresponding product distribution $x^{(i)}$ and indicates to the ellipsoid algorithm that $(x^{(i)}U^T)y \leq -1$ is violated by $y^{(i)}$. The ellipsoid algorithm will stop after a polynomial number of steps and determine that the program is infeasible. Let $X$ be the matrix whose rows are the generated product distributions $x^{(1)}, \ldots, x^{(L)}$.

Consider the linear program

$$[XU^T]y \leq -1, y \geq 0, \qquad (D')$$

and observe that the rows of $[XU^T]y \leq -1$ are the cuts generated by the ellipsoid method. If we apply the same ellipsoid method to $(D')$ and use a separation oracle that returns the cut $x^{(i)} U^T y \leq -1$ given query $y^{(i)}$, the ellipsoid algorithm would go through the same sequence of queries $y^{(i)}$ and cutting planes $x^{(i)} U^T y \leq -1$ and return infeasible. Presuming that numerical problems do not arise,[2] we will find that $(D')$ is infeasible. This implies that its dual $[UX^T]\alpha \geq 0$, $\alpha \geq 0$ is unbounded and has polynomial size, and thus can be solved for a nonzero feasible $\alpha$. We can thus scale $\alpha$ to obtain a probability distribution. We then observe that $X^T \alpha$ satisfies the incentive constraints (1) and the probability distribution constraints (2) and is therefore a correlated equilibrium. The distribution $X^T \alpha$ is the mixture of product distributions $x^{(1)}, \ldots, x^{(L)}$ with weights $\alpha$, and thus can be represented in polynomial space and can be efficiently sampled from.

One issue remains. Although the matrix $XU^T$ is polynomial sized, computing it using matrix multiplication would involve an exponential number of operations. On the other hand, entries of $XU^T$ are differences between expected utilities that arise under product distributions. Since we have assumed that the game representation admits a polynomial-time algorithm for computing such expected utilities, $XU^T$ can be computed in polynomial time.

**Lemma 3.2** ([Papadimitriou, 2005; Papadimitriou & Roughgarden, 2008]). *There exists an algorithm that given a game representation with polynomial type and polynomial expectation property, and given an arbitrary product distribution $x$, computes $xU^T$ in polynomial time. As a result, $XU^T$ can be computed in polynomial time.*

## 4 Our Algorithm

In this section we present our modification of the Ellipsoid Against Hope algorithm, and prove that it computes exact CE. There are two key differences between our approach and the original algorithm for computing approximate CE.

1. Our modified separation oracle produces pure-strategy-profile cuts;

2. The algorithm is simplified, no longer requiring a special mechanism to deal with numerical issues (because pure-strategy-profile cuts can be represented directly as rows of $(D)$'s constraint matrix).

---

[2]Since each row of $(D')$'s constraint matrix $XU^T$ may require more bits to represent than any row of the constraint matrix $U^T$ for $(D)$, running the ellipsoid algorithm on $(D')$ with the original bounding ball and volume lower bound for $(D)$ would not be sound, and as a result $(D')$ is not guaranteed to be infeasible. Indeed, Stein *et al.* [2010] showed that when running the algorithm on their symmetric game example, $(D')$ would remain feasible, and thus the output of the algorithm would not be an exact CE. Furthermore, since the only CE of that game that is a mixture of symmetric product distributions is irrational, there is no way to resolve this issue without breaking at least one of the symmetry and product distribution properties of the Ellipsoid Against Hope algorithm. For more on these issues and possible ways to address them, please see Papadimitriou and Roughgarden [2008] and Stein, Parrilo & Ozdaglar [2010].

## 4.1 The Purified Separation Oracle

We start with a "purified" version of Lemma 3.1.

**Lemma 4.1.** *Given any dual vector $y \geq 0$, there exists a pure strategy profile $s$ such that $(U_s)^T y \geq 0$.*

*Proof.* Recall that Lemma 3.1 states that given dual vector $y \geq 0$, a product distribution $x$ can be computed in polynomial time such that $xU^T y = 0$. Since $x[U^T y]$ is a convex combination of the entries of the vector $U^T y$, there must exist some nonnegative entry of $U^T y$. In other words, there exists a pure strategy profile $s$ such that $(U_s)^T y \geq xU^T y = 0$. □

The proof of Lemma 4.1 is a straightforward application of the probabilistic method: since $xU^T y$ is the expected value of $(U_s)^T y$ under distribution $x$, which we denote $E_{s \sim x}[(U_s)^T y]$, the nonnegativity of this expectation implies the existence of some $s$ such that $(U_s)^T y \geq 0$. Like many other probabilistic proofs, this proof is not efficiently constructive; note that there are an exponential number of possible pure strategy profiles.

It turns out that for game representations with polynomial type and satisfying the polynomial expectation property, an appropriate $s$ can indeed be identified in polynomial time. Our approach can be seen as derandomizing the probabilistic proof using the method of conditional probabilities [Erdos & Selfridge, 1973; Spencer, 1994; Raghavan, 1988]. At a high level, for each player $p$ our algorithm picks a pure strategy $s_p$, such that the conditional expectation of $(U_s)^T y$ given the choices so far remains nonnegative. This requires us to compute the conditional expectations, but this can be done efficiently using the expected utility subroutine guaranteed by the polynomial expectation property.

**Lemma 4.2.** *There exists a polynomial-time algorithm that given*

- *an instance of a game in a representation satisfying polynomial type and the polynomial expectation property,*
- *a polynomial-time subroutine for computing expected utility under any product distribution (as guaranteed by the polynomial expectation property), and*
- *a dual vector $y \geq 0$,*

*finds a pure strategy profile $s \in S$ such that $(U_s)^T y \geq 0$.*

*Proof.* Given a product distribution $x$, let $x_{(p \to s_p)}$ be the product distribution in which player $p$ plays $s_p$ and all other players play according to $x$. Since $x$ is a product distribution, $x_{(p \to s_p)} U^T y$ is the conditional expectation of $(U_s)^T y$ given that $p$ plays $s_p$, and furthermore we have for any $p$,

$$xU^T y = \sum_{s_p} \left[ x_{(p \to s_p)} U^T y \right] x^p_{s_p}. \qquad (3)$$

**Algorithm 1** Computes a pure strategy profile $s$ such that $(U_s)^T y \geq 0$.

1. Given $y \geq 0$, identify a product distribution $x$ satisfying $xU^T y = 0$, using the algorithm described in Lemma 3.1.

2. For each player $p$,

   (a) iterate through actions $s_p \in S_p$, and compute $x_{(p \to s_p)} U^T$ using the algorithm described in Lemma 3.2, until we find an action $s_p^* \in S_p$ such that $\left[ x_{(p \to s_p^*)} U^T \right] y \geq 0$.

   (b) set $x$ to be $x_{(p \to s_p^*)}$.

3. The resulting $x$ corresponds to a pure strategy profile $s$. Output $s$.

---

Since $x^p$ is a distribution, the right hand side of (3) is a convex combination and thus there must exist an action $s_p \in S_p$ such that $x_{(p \to s_p)} U^T y \geq xU^T y \geq 0$. Since $x_{(p \to s_p)}$ is a product distribution, this process can be repeated for each player to yield a pure strategy profile $s$ such that $(U_s)^T y \geq xU^T y \geq 0$. This is formalized in Algorithm 1.

We now consider the running time of Algorithm 1. Due to the polynomial expectation property, the algorithm described in Lemma 3.2 is polynomial, which implies that in Step 2a, for each $s_p \in S_p$, $x_{(p \to s_p)} U^T$ can be computed in polynomial time. Since Step 2a requires at most $|S_p|$ such computations, and since polynomial type implies that $n$ and $|S_p|$ are polynomial in the input size, the algorithm runs in polynomial time. □

A straightforward corollary is the following:

**Corollary 4.3.** *Algorithm 1 can be used as a separation oracle for the dual LP (D) in the Ellipsoid Against Hope algorithm: for each query point $y$, the oracle computes the corresponding pure-strategy profile $s$ according to Algorithm 1 and returns the half space $(U_s)^T y \leq -1$. We call this the Purified Separation Oracle. This separation oracle has the following properties:*

- *Each returned half space is one of the constraints of (D).*

- *Since Algorithm 1 iterates through the players sequentially, the generated pure-strategy profiles can be asymmetric even for symmetric games and symmetric $y$.*

- *Since a pure-strategy profile is a special case of a product distribution, the resulting pure-strategy profile $s$ also satisfies Lemma 3.1, with $x$ being the unit vector corresponding to $s$.*

## 4.2 The Simplified Ellipsoid Against Hope Algorithm

We now modify the Ellipsoid Against Hope Algorithm by replacing the Product Separation Oracle with our Purified Separation Oracle. The rows of $X$ in $(D')$ become unit vectors corresponding to the pure-strategy profiles generated by the oracle. Thus, we can write $(D')$ as

$$(U')^T y \leq -1, y \geq 0, \qquad (D'')$$

where the matrix $U' \equiv UX^T$ consists of the columns $U_{s^{(i)}}$ that correspond to pure-strategy profiles $s^{(i)}$ generated by the separation oracle. Note that each constraint of $(D'')$ is also one of the constraints of $(D)$, and as a result neither the maximum value of the coefficients nor the right-hand sides of $(D'')$ are greater than in $(D)$. Therefore, a starting ball and volume lower bound that are valid for a run of the ellipsoid method on $(D)$ would also be valid for $(D'')$. We thus avoid the precision issues faced by the Ellipsoid Against Hope algorithm, and it is sufficient to use standard values for the initial radius and volume lower bound (and standard perturbation methods for dealing with non-full-dimensional solutions). The resulting CE is a mixture over a polynomial number of pure strategy profiles. We can make a further conceptual simplification of the algorithm: instead of using $X$ as in the Ellipsoid Against Hope algorithm, we can directly treat the generated pure-strategy profiles as columns of $U$, and use $U'$ in place of $UX^T$.

We now formally state and prove our result. Note that although we only briefly discussed the way numerical issues are addressed in the original Ellipsoid Against Hope algorithm in Section 3, we need to go into detail about how our algorithm ensures its own numerical accuracy. Nevertheless that task is comparatively easy, as it is sufficient for us to directly apply standard techniques from the theory of the ellipsoid method. Our analysis makes use of the following lemma from [Grötschel et al., 1988].

**Lemma 4.4** (Lemma 6.2.6, [Grötschel et al., 1988]). *Let $P = \{y \in \mathbb{R}^N | Ay \leq b\}$ be a full-dimensional polyhedron defined by the system of inequalities, with the encoding length of each inequality at most $\varphi$. Then $P$ contains a ball with radius $2^{-7N^3\varphi}$. Moreover, this ball is contained in the ball with radius $2^{5N^2\varphi}$ centered at 0.*

We note that the only restriction on $P$ is full dimensionality; we do not need to assume that $P$ is bounded, or that $A$ has full row rank.

**Theorem 4.5.** *Given a game representation with polynomial type and polynomial expectation property, Algorithm 2 computes an exact and rational CE with support size at most $1 + \sum_p |S_p|(|S_p| - 1)$ in polynomial time.*

*Proof.* We begin by proving the correctness of the algorithm. First, we will show that the ellipsoid method in Step 1 is a valid run for $(D)$, which certifies that

**Algorithm 2** Computes an exact rational CE given a game representation satisfying polynomial type and the polynomial expectation property.

1. Apply the ellipsoid method to $(D)$, using the Purified Separation Oracle, a starting ball with radius of $R = u^{5N^3}$ centered at 0, and stopping when the volume of the ellipsoid is below $v = \alpha_N u^{-7N^5}$, where $\alpha_N$ is the volume of the $N$-dimensional unit ball.

2. Form the matrix $U'$ whose columns are $U_{s^{(1)}}, \ldots, U_{s^{(L)}}$ generated by the separation oracle during the run of the ellipsoid method.

3. Compute a basic feasible solution $x'$ of the linear feasibility program

$$U'x' \geq 0, x' \geq 0, \mathbf{1}^T x' = 1, \qquad (P^*)$$

by applying the ellipsoid method on the explicitly represented $(P^*)$ and recovering a basis using, e.g., Algorithm 4.2 of [Dantzig & Thapa, 2003].

4. Output $x'$ and $s^{(1)}, \ldots, s^{(L)}$, interpreted as a distribution over pure-strategy profiles $s^{(1)}, \ldots, s^{(L)}$ with probabilities $x'$.

the feasible set of $(D)$ is either empty or not full dimensional.[3] Suppose the contrary, i.e., the feasible set of $(D)$ is feasible and full dimensional. Since the encoding length of each constraint of $(D)$ is at most $N \log_2 u$, then by Lemma 4.4, the feasible set must contain a ball with radius $u^{-7N^4}$, and thus volume $\alpha_N u^{-7N^5}$, and furthermore this ball must be contained in the ball with radius $u^{5N^3}$ centered at 0, which is the initial ball of our ellipsoid method in Step 1. Since at the end of Step 1 the ellipsoid method certifies that the intersection of the initial ball and the feasible set has volume less than $v = \alpha_N u^{-7N^5}$, we reach a contradiction and therefore either the LP $(D)$ must be infeasible or the feasible set must not be full dimensional. Since the largest magnitude of the coefficients in $(D'')$ is also $u$, Step 1 is also a valid run for $(D'')$ and therefore either $(D'')$ must be infeasible or the feasible set of $(D'')$ must not be full dimensional.

Of course a non-full-dimensional feasible set is not sufficient for our purpose; we now perturb $(D'')$ to get an infeasible LP. Fix $\rho > 1$. Perturbing the constraints $(U')^T y \leq -1$ of $(D'')$ by multiplying the RHS by $\rho$, we get the following LP:

$$\min 0 \qquad (4)$$
$$(U')^T y \leq -\rho \mathbf{1}$$
$$y \geq 0.$$

---
[3]Since the ellipsoid method relies on shrinking the volume of the candidate set, it is not able to distinguish between non-full-dimensional feasible sets and infeasibility. We overcome this by perturbing the LP after the ellipsoid method has been applied; another method perturbs the LP in advance to ensure the feasible set is either empty or full dimensional.

We claim that (4) is infeasible. Suppose otherwise: then there exists a $y \in \mathbb{R}^N$ such that $y \geq 0$ and $(U')^T y \leq -\rho \mathbf{1}$. Let $y' \in \mathbb{R}^N$ be a vector such that $0 \leq y'_j - y_j \leq \frac{\rho-1}{Nu}$ for all $j$. Then $y' \geq 0$, and each component $s$ of $U'^T y'$ satisfies

$$(U'_s)^T y \leq (U'_s)^T y + \frac{\rho-1}{Nu} \sum_j |U'^j_s|$$
$$\leq -\rho + \rho - 1$$
$$\leq -1.$$

Thus, any such $y'$ is feasible for $(D'')$. However, the set of all such vectors $y'$ is a full-dimensional cube. This contradicts the fact that $(D'')$ is either infeasible or not full dimensional, and therefore (4) is infeasible. This means that (4)'s dual

$$\max \rho \mathbf{1}^T x' \qquad (5)$$
$$U' x' \geq 0$$
$$x' \geq 0$$

is unbounded (since it is feasible, e.g. $x' = 0$). Then a nonzero feasible vector $x'$ is (after normalization) a distribution over the pure strategy profiles corresponding to columns of $U'$. Treating it as a sparse representation of a correlated distribution $x$, it satisfies the feasibility program for CE and is therefore an exact CE.

This CE is exact but its support could be greater than $1 + \sum_p |S_p|(|S_p|-1)$ (although as we argue below it is still polynomial). To get a CE with the required support size, we notice that since (5) is unbounded, a feasible solution of the bounded linear feasibility program $(P^*)$ is a CE. Note that $(P^*)$ has the same set of rows constraints as the feasibility program for CE defined by (1) and (2), and that for each player $p$ and action $i \in S_p$, the incentive constraint $(p, i, i)$ corresponds to deviating from action $i$ to itself and is therefore redundant. Thus the number of bounding constraints of $(P^*)$ is at most $1 + \sum_p |S_p|(|S_p|-1)$ and therefore a basic feasible solution $x'$ of $(P^*)$ will have the required support size. Since the coefficients and right-hand sides of $(P^*)$ are rational, then (by e.g. Lemma 6.2.4 of [Grötschel et al., 1988]) its basic feasible solution $x'$ is also rational and can be represented using at most $4N^3 u$ bits.

We now consider the running time of the algorithm. Since Step 1 is a standard run of the ellipsoid method, it terminates in a polynomial number of iterations. For example if we use the ellipsoid algorithm presented in Theorem 3.2.1 of [Grötschel et al., 1988], then by Lemma 3.2.10 of [Grötschel et al., 1988] the ratio between volumes of successive ellipsoids $\text{vol}(E_{k+1})/\text{vol}(E_k) \leq e^{-1/(5N)}$. With the volume of the initial ellipsoid at most $\alpha_N R^N$ and stopping when vol-ume is below $v$, the number of iterations $L$ is at most

$$\begin{aligned} &5N\left[\ln(\alpha_N R^N) - \ln v\right] \\ &= 5N\left[5N^4 \ln u + 7N^5 \ln u\right] \\ &= O(N^6 \ln u), \end{aligned}$$

which is polynomial in the input size since $N$ is polynomial. Since each call to the separation oracle takes polynomial time by Lemma 4.2, Step 1 takes polynomial time. $L$ being polynomial also ensures that $(P^*)$ has polynomial size, and thus a basic feasible solution can be found in polynomial time. □

We note that the estimates on $R$ and $v$ (and thus $L$) can be improved, but our main goal here is to prove that the running time of our algorithm is polynomial.

The reader may wonder how our algorithm would deal with Stein *et al.* [2010]'s counterexample, a symmetric game in which the only CE that is a convex combination of symmetric product distributions has irrational probabilities. Since we have proved that our algorithm computes a rational CE as a convex combination of product distributions, it must violate the symmetry property. Indeed as we discussed in Section 4.1, our Purified Separation Oracle can return asymmetric cuts for symmetric games and symmetric queries, and thus for this game it must return at least one asymmetric cut during the algorithm.

## 5 Conclusion

We have proposed a polynomial-time algorithm, a variant of Papadimitriou and Roughgarden's Ellipsoid Against Hope approach, for computing an exact CE given a game representation with polynomial type and polynomial expectation property. A key component of our approach is a derandomization of Papadimitriou and Roughgarden's separation oracle using the method of conditional probabilities, yielding a polynomial-time separation oracle that outputs cuts corresponding to pure-strategy profiles. Our approach is then spared from dealing with the numerical precision issues that were a major focus of previous approaches, and the algorithm is considerably simplified as a result. Furthermore, the correlated equilibria returned by our algorithm have polynomial-sized supports. We expect these properties of our algorithm to be independently interesting, beyond its usefulness in resolving the recent uncertainty about the computational complexity of identifying exact CE.

Our algorithm has additional practical benefits: the resulting cutting planes are deeper cuts than those produced by the original oracle, resulting in a smaller number of iterations required to reach convergence, albeit at the cost of more work per iteration. It is also possible to return cuts corresponding to pure strategy profiles with (e.g.) good social welfare, yielding a heuristic method for generating correlated equilibria with good social welfare; we do note, however, that finding a CE with optimal social welfare is generally NP-hard for many game representations [Papadimitriou & Roughgarden, 2005].